\documentclass[final,authoryear,5p,times,twocolumn]{elsarticle}

\usepackage{graphics}
\usepackage{amssymb}
\usepackage{epigraph}

\usepackage{lineno}

\journal{Planetary and Space Science}

\begin{document}

\twocolumn[
\section*{\Large {\mdseries Editorial}}
\section*{\LARGE {\sc Cosmic Dust \MakeUppercase{\romannumeral 8}}}
\begin{flushleft}
\hrulefill
\end{flushleft}
\vspace{0.3in}
]

{\sl \epigraph{For dust you are and to dust you will return}{Genesis 3:19}}

\section{Interdisciplinary research on cosmic dust}

These days, once taking a look at cosmic environments in various spectral ranges, one can catch a glimpse of the dusty world in diverse shapes of appearance. 
Apart from its elegancy and beauty in the appearance, cosmic dust has a great influence on a variety of physical processes and complex chemical evolution in different epochs and sites of the Universe. 
In widely accepted scenarios of the cosmos such as the evolution of stars and the formation of planetary systems, dust as well as gas is the driving force. 
Even though it is frustrating to find that cosmic dust obscures vision in the Universe, it is fascinating to study the dust as the building blocks of planets and life.
Our planet Earth, for example, is simply made by the accumulation of dust grains initially condensed in the disk around our new-born Sun about 4.6 billion years ago.
On the one hand, silicates and metals, major constituents of cosmic dust, were the most important and predominant contributors to the development of Earth's crust, mantle, and core. 
On the other hand, organic-rich carbonaceous material, another major constituent of the dust, may have played a vital role in the origin and evolution of life on the early Earth.
The surface of dust grains provides a favorable site on which gaseous species can freeze out and eventually react between each other resulting in a rich zoo of organic compounds.  
Owing to the overwhelming chemical and structural multiplicity of such organic materials, cosmic dust may be the main disseminator of life across the Universe.

When grazing up at a night sky, we may find it filled with a wide variety of unique and interesting dusty phenomena at stars, nebulae, clouds, and planetary systems.
However, no matter how long we look into such a phenomenon, it is only a blink of an eye on a timescale of the entire life of cosmic dust. 
To understand the whole life cycle of the dust, we need to unravel the deep mystery of dust coming out of evolved stars, traveling between different phases of the interstellar medium, dying out at protostars, reviving around young stars, and finally seeding into planetary systems.  
Solving this mystery is a serious challenge whereof scientists from all over the world are working continuously and sharing a common interest. 
This is the ultimate goal, for which observers, theorists, and experimentalists have to work hand in hand with each other and need to combine their activities and studies for the optimal success. 
The annual Cosmic Dust meeting provides an ideal platform for scientists tackling cosmic dust problems in a wide range of disciplines to meet each other and intensively discuss their recent results and developments in this field. 
This meeting started in 2006 as a session called ``Cosmic Dust'' of the 3rd AOGS (Asia-Oceania Geosciences Society) annual meeting in Singapore, but detached itself from the AOGS since 2012. 
We have successfully kept organizing this series of Cosmic Dust meetings in a relaxing atmosphere with an excellent choice of speakers until today owing to our extraordinary enthusiasms about the development of cosmic dust research.
The Cosmic Dust meeting features the coziness, ending up with grandiose contributions by a great number of top scientists from all around the world, which made the meeting a striking success. 

\begin{figure*}[t]
 \begin{center}
  \includegraphics[width=180mm]{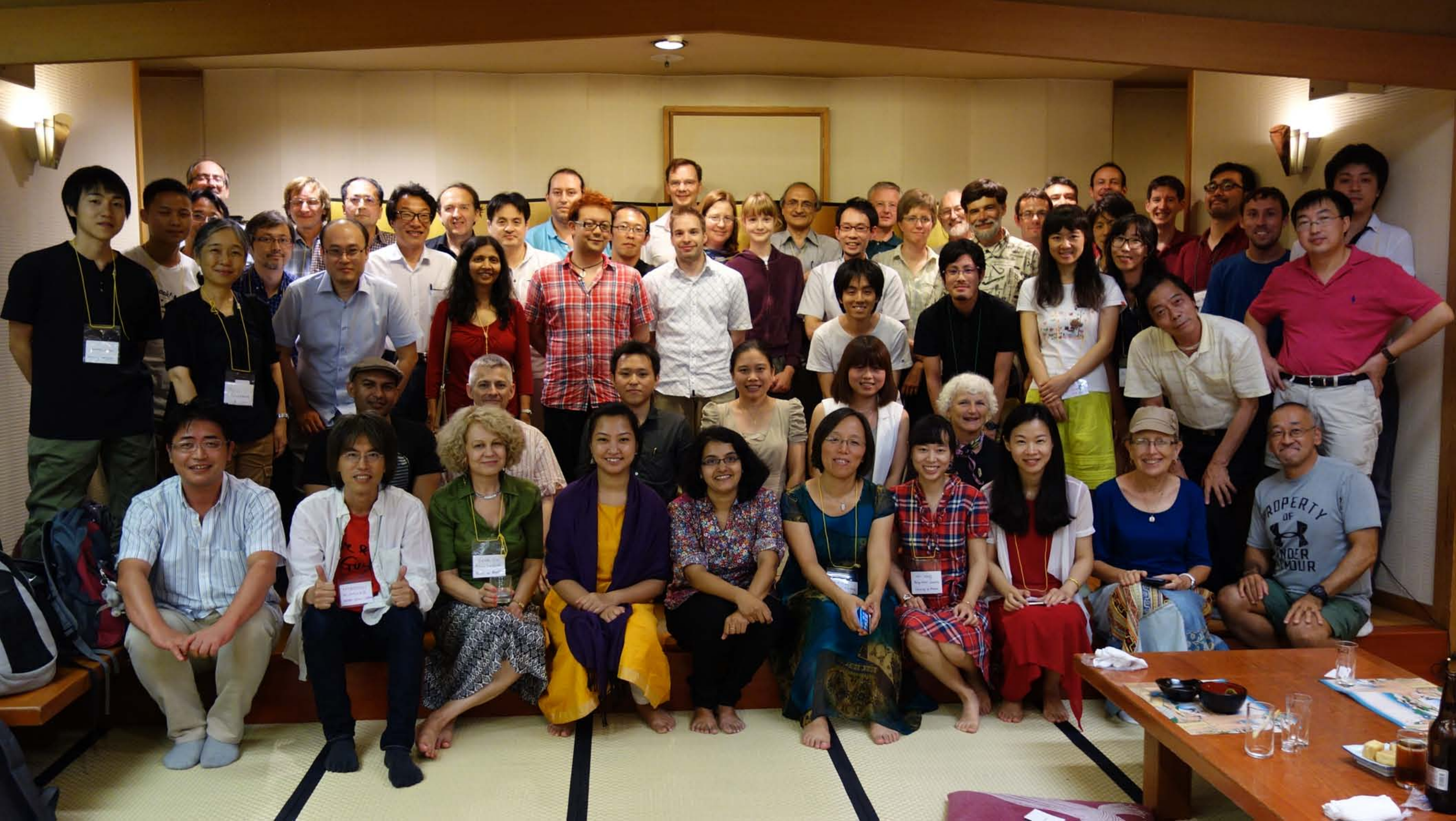}
 \end{center}
 \caption{A group picture of participants to {\sc Cosmic Dust \MakeUppercase{\romannumeral 8}}; (In no particularly order) 
 M. Fulle, M. Hilchenbach, H. Hirashita, R. Botet, F. Galliano, P. Woitke, J. Murthy, H. Kr\"{u}ger, M. Podolak, C. Pinte, D. Lazzati, G. Kennedy, M.S. Bentley, J. Li, H. Senshu, T. Ootsubo, H. Ding, K. Bekki, T. Nozawa, C. Helling, T. Hirai, V. Sterken, G. Vidali, T. Kondo, H. Kobayashi, T. Kokusho, A. M. Nakamura, J. Pyo, V.P. Kulkarni, M. Hammonds, C.T. Tibbs, S. Hamano, R. Tazaki, Y. Xie, J.Y. Seok, B. Mills, M. Kiuchi, Z. Wahhaj, G. Rouill\'{e}, T. Shimonishi, X. Yang, X. Chen, T. Onaka, A. Li, K.Wada, H. Kimura, L. Kolokolova, M. Buragohain, Shalima P., B. Jiang, S. Wang, L. Zhou, and H. Chihara.}
 \label{fig:one}
\end{figure*}

The 8th meeting on Cosmic Dust (hereafter {\sc Cosmic Dust \MakeUppercase{\romannumeral 8}}) was held at the Tokyo Skytree Town Campus of Chiba Institute of Technology, Japan, from August 17 to 21, 2015. 
This special issue in Planetary and Space Science (PSS) is primarily intended for the papers presented at {\sc Cosmic Dust \MakeUppercase{\romannumeral 8}}.
Nevertheless, we are pleased that this special issue also contains papers beyond the presentations at the meeting such as \citet{hooks-et-al2016}, \citet{kolokolova-et-al2016}, \citet{marboeuf-et-al2016}, \citet{mattsson2016}, and \citet{podolak-et-al2016}.
The full scientific program of {\sc Cosmic Dust \MakeUppercase{\romannumeral 8}} is available at the Cosmic Dust website (https://www.cps-jp.org/\~{}dust/Program\_VIII.html) with links to the abstracts of all presentations.
Needless to say, invited talks, contributed talks and posters\footnote{\citet{xie-et-al2016} and \citet{tazaki-et-al2015} received the Best Poster Award of {\sc Cosmic Dust \MakeUppercase{\romannumeral 8}}, on the basis of judgements made by the invited speakers, Varsha P. Kulkarni (University of South Carolina, USA), Gianfranco Vidali (Syracuse University, USA), Christophe Pinte (CNRS/INSU, France), Grant Kennedy (University of Cambridge, UK), Charles M. Telesco (University of Florida, USA), Harald Kr\"{u}ger (Max-Planck-Institut f\"{u}r Sonnensystemforschung, Germany), Martin Hilchenbach (Max-Planck-Institut f\"{u}r Sonnensystemforschung, Germany), Marco Fulle (INAF Trieste, Italy), Mark S. Bentley (Institut f\"{u}r Weltraumforschung Graz, Austria), Andrew Westphal (University of California at Berkeley, USA), and Hongbin Ding (Dalian University of Technology, China).}, as well as on-site discussions were dedicated to many different aspects of cosmic dust.
In Section~\ref{review}, therefore, we intend to provide a critical overview of the major scientific issues that were addressed in {\sc Cosmic Dust \MakeUppercase{\romannumeral 8}}, so that the reader could grasp the most recent hot topics of cosmic dust research.
The main social events of the meeting were a half-day excursion to the Edo-Tokyo Open Air Architectural Museum in Koganei Park and a banquet held at a traditional Japanese kaiseki restaurant, named ``Kappou Sansyu-ya'', established in 1889.
Fig.~\ref{fig:one} is a group shot of the participants in a Japanese-style tatami-floored hall of the restaurant taken immediately after the banquet.
Without a shadow of doubt, the Local Organizing Committee (LOC) led by Koji Wada (PERC/Chitech, Japan) did an excellent job in organizing the meeting along the rationale behind it.
This editorial will end with our remarks on {\sc Cosmic Dust \MakeUppercase{\romannumeral 8}} in Section~\ref{perspectives}, along with our perspectives on the development of cosmic dust research.

\section{The contents of {\sc \bf Cosmic Dust \MakeUppercase{\romannumeral 8}}}
\label{review}

\subsection{Galaxies}

The characterization of dust in distant galaxies is of vital importance for a better understanding of galaxy evolution and the cosmic expansion rate at earlier epochs.
Varsha P. Kulkarni presented observational evidence that the composition of dust in distant galaxies may be systematically different from that in our Galaxy and nearby galaxies \citep{kulkarni-et-al2016}.
The most notable difference is the predominance of crystalline silicates in some distant galaxies, because amorphous silicates are dominant in nearby galaxies.
The high crystallinity might be associated with low cosmic-ray fluxes because exposure to cosmic rays transforms a crystalline phase into an amorphous phase \citep[e.g.,][]{kraetschmer-huffman1979,jaeger-et-al2003}.
The observed spectral features of crystalline silicates were best fitted with laboratory profiles of moderately iron-rich crystalline silicates \citep{aller-et-al2012,aller-et-al2014}.
We should point out that if crystalline silicates in distant galaxies would formed by condensation of gas and did not undergo amorphization, then the silicates tend to be magnesium-rich, iron-poor crystals \citep{molster-kemper2005}.
In contrast, laboratory experiments on annealing of amorphous silicates with embedded metal have shown that the annealing transforms the silicates into moderately iron-rich olivine and pyroxene \citep{brownlee-et-al2005}.
{\sl Accordingly, we expect that moderately iron-rich crystalline silicates detected in distant galaxies arose from annealing of amorphous silicates, rather than condensation of gas.}

Other papers covered the following topics: a study on dust properties at high redshifts from a modeling of dust extinction in gamma-ray bursts by Tayyaba Zafar \citep{zafar2016}; simplification of dust size distribution for a modeling of dust enrichments in galaxies by Hiroyuki Hirashita \citep{hirashita-et-al2016}; computer simulations of galaxy formation and evolution with dust physics by Kenji Bekki; detailed analyses of AKARI, 2MASS, and WISE data on spectral energy distributions of early-type galaxies by Takuma Kokusho; a modeling of silicate dust grains in the circumnuclear torus of active galactic nuclei by Yanxia Xie \citep{xie-et-al2016}; a modeling of intergalactic extinction based on the depletion of dust-forming elements by Bradley Mills.

\subsection{Evolved stars, supernovae, and star-forming regions}

Owing to the third-largest abundance in the local universe after hydrogen and helium, oxygen plays an important role in the chemistry of gas and dust.
Dust particles incorporate oxygen in a variety of forms such as ices, silicates, and organic-rich carbonaceous material, depending on their environments \citep[e.g.,][]{whittet2010}.
Gianfranco Vidali reported laboratory experiments on the chemical reactions of oxygen at grain surfaces carried out in his group as well as theoretical simulations on the reactions in photo-dissociation regions and star-forming regions \citep{vidali2015}.
All recent experimental studies indicate that the binding energy of oxygen is approximately double the value previously adopted in astrochemical models \citep{ward-price2011,kimber-et-al2014,he-et-al2014,he-et-al2015,minissale-et-al2016}.
A higher binding energy of O atoms leads to an increase in the residence time of O atoms on grain surfaces and enhances the abundance of H$_2$O ice \citep[see][]{he-et-al2015}. 
Very recently, \citet{wang-et-al2015} and \citet{poteet-et-al2015} have postulated a population of very large, micrometer-sized H$_2$O ice grains in the diffuse interstellar medium to account for the observed mid-infrared extinction as a major repository for O atoms depleted from the gas phase.
However, micrometer-sized H$_2$O ice grains are not exactly a requisite for the mid-infrared extinction, because micrometer-sized amorphous carbon grains could also help to reproduce the observed mid-infrared extinction \citep{wang-et-al2014}.
Furthermore, size degradation of large grains by photodesorption, shattering, and sputtering would continuously replenish small grains, but small H$_2$O ice grains are absent in the diffuse interstellar medium \citep[cf.][]{whittet-et-al1988}.
The processing of H$_2$O ice by exposure to ultraviolet radiation and cosmic rays also results in the formation of organic refractory material as well as polycyclic aromatic hydrocarbons (PAHs) \citep{greenberg-et-al2000}.
Therefore, organic refractory material could be the main carrier of O atoms in the dust phase, although spectroscopic observations have not provided conclusive evidence \citep{jenkins2009,whittet2010}.
{\sl Consequently, the new determination of oxygen binding energy alone has not yet given a solution to the problem of the ``missing'' oxygen atoms, which has been a matter of debate for several decades.}

Other papers covered the following topics: a numerical model of dust nucleation and growth in young supernova remnants by Davide Lazzati; a nucleation and growth model of small carbon grains based on the ab-initio density functional theory by Christopher Mauney \citep{mauney-lazzati2016}; a modeling of the reddening law observed for Type Ia supernovae by Takaya Nozawa \citep{nozawa2016}; a modeling of the color excesses toward the Type Ia supernova SN 2014J in the starburst galaxy M82 by Jun Li; a revision of the infrared extinction law based on the APOGEE spectroscopic survey by Biwei Jiang; a model constraint on the abundance of small spinning grains in dense cores based on Planck centimeter observations by Christopher T. Tibbs.

\subsection{Protoplanetary disks}

The collisional growth of dust particles in protoplanetary disks is currently considered a plausible scenario for the formation of planetesimals and planets \citep{weidenschilling-et-al1989}.
Christophe Pinte indicated that a classical theory of planetesimal formation has shortcomings in the mechanism of collisional grain growth, which has not yet been resolved by observations \citep{pinte2015}.
Meter-sized dust aggregates collide so fast that they cannot stick to grow further, but they suffer from fragmentation,  the so-called fragmentation barrier \citep[e.g.,][]{dullemond2013}. 
However, \citet{wada-et-al2009} have shown that dust aggregates consisting of submicrometer-sized grains can go beyond the fragmentation barrier, if they are composed of ices.
Subsequently, the reassessment of surface energy for amorphous silica revealed that the stickiness of silicate dust aggregates has been underestimated by one to two orders of magnitude \citep{kimura-et-al2015a}.
Furthermore, the sticking efficiency of dust aggregates can be greatly enhanced, if submicrometer-sized mineral grains are encased in organic refractory materials \citep{flynn-et-al2013,kimura-et-al2015a}.
Indeed, the chondritic porous subset of interplanetary dust particles, which is most likely of cometary origin, is an aggregate of submicrometer-sized grains whose outer layers are covered by organic-rich carbonaceous materials \citep{keller-et-al2000,flynn-et-al2013}.
{\sl Therefore, the presence of organic refractory materials may be an important aid to overcome the fragmentation barrier and thus plays a vital role in the formation of planetesimals in protoplanetary disks.}

Other papers covered the following topics: numerical evaluation of ejecta mass produced by collisions of single grains onto dust aggregates by Koji Wada; numerical simulations on large-scale radial transport and mixing of centimeter-sized grains in a marginally gravitationally unstable disk by Morris Podolak; a development of a holistic modeling approach to dust and gas in protoplanetary disks by Peter Woitke; a development of a model for dust formation in atmospheres of planetary objects by Christiane Helling; an application of the Rayleigh-Gans-Debye theory to computations of phase function for fractal dust aggregates by Ryo Tazaki \citep{tazaki-et-al2015}; numerical simulations of polarimetric disk images in the mid-infrared wavelengths by Han Zhang.

\subsection{Debris disks}

Debris disks have been detected and imaged around a number of main-sequence stars since the discoveries of the disks around Vega by \citet{aumann-et-al1984} and $\beta$ Pictoris by \citet{smith-terrile1984}.
Early studies on debris disks have been limited to the outer part of the disks, since only the population of cold dust originating from the exo-Kuiper belts was spatially resolved.
Recent infrared interferometric observations of debris disks have provided an opportunity to explore the inner part of the disks, namely, the exo-zodiacal light from warm dust, or even the innermost part of the disk, the exo-F-corona of hot dust.
Near-infrared observations of debris disks around nearby main-sequence stars have revealed no clear correlation between excess emission from hot dust grains and the presence of cold dust populations \citep{ertel-et-al2014}.
Because all the proposed mechanisms for supplying the great population of hot dust in the vicinity of stars are by no means conclusive, Grant Kennedy stated that the presence of the hot dust population is a total mystery \citep{kennedy2015}.
It might be worthwhile to recognize that near-infrared excess emission observed in the solar F-corona is a similar enigmatic phenomenon.
The excess emission of the solar F-corona has been detected around $4\,R_\odot$ from the Sun until 1983 in the H and J bands and until 1973 in the K and L bands, but its origin is still an open question \citep{kimura-mann1998}.
Here we cannot help but wonder whether the hot dust populations could be replenished by the super-catastrophic disruption of sub-kilometer to kilometer sized planetesimals in the vicinity of the star \citep[cf.][]{granvik-et-al2016}.
{\sl If the mechanism to supply the population of hot dust in debris disks would be identified, then it might give a clue about the origin of the near-infrared excess emission in the solar F-corona, and vice versa.}

Other papers covered the following topics: early results of the mid-IR multi-mode facility camera CanariCam for determining dust properties in young circumstellar disks by Charles M. Telesco; a modeling of dust and PAHs in the debris disk around HD 34700 based on the spectral energy distribution of the disk by Ji Yeon Seok; an improvement on the signal to noise ratio in the direct imaging of the HD 32297 debris disk by Zahed Wahhaj; numerical simulation on the coagulation and fragmentation of planetesimals in debris disks by Hiroshi Kobayashi.

\subsection{Solar System}

After a decade-long journey to the Jupiter-family comet 67P/Churyumov-Gerasimenko (67P/C-G), Rosetta is now in the midst of in-situ measurements with various instruments onboard the Rosetta orbiter and the Philae lander \citep{taylor-et-al2015}.
Harald Kr\"{u}ger was the first speaker in a series of the invited talks dedicated to successful results of the measurements acquired by Rosetta's dust instruments \citep{krueger-et-al2015}.
He introduced the Dust Impact Monitor (DIM) and the Cometary Acoustic Sounding Surface Experiment (CASSE), which are parts of the Surface Electric Sounding and Acoustic Monitoring Experiment (SESAME) onboard Philae, and the Cometary Sampling and Composition (COSAC) experiment, with particular emphasis on the DIM.
The DIM measurements were not successful before the separation from the orbiter and after landing on the nucleus due to unfavorable positions of the instrument.
However, during the short period of Philae's descent to the nucleus, the DIM detected one and only millimeter-sized dust particle at an altitude of 2.4 km from the surface.
Laboratory calibration experiments show that the measured signal amplitude and long contact duration are consistent with an impact of a porous particle with a density $\rho = 250\,\mathrm{kg\,m^{-3}}$.
We are aware that millimeter-sized primordial dust aggregates have a much smaller density of $\rho \sim 0.1\,\mathrm{kg\,m^{-3}}$ in protoplanetary disks based on a model of particle growth by coagulation \citep{weidenschilling-et-al1989}.
Although 67P/C-G's millimeter-sized particles may have a broad range of densities, the DIM's measurement already provides strong evidence that millimeter-sized particles could have a density of $\rho \gg 0.1\,\mathrm{kg\,m^{-3}}$.
{\sl Therefore, we may conclude that dust particles in comets have experienced intense compaction at some stage such as a certain period of the formation of comet nuclei or the formation of a dust mantle on their surfaces.}

Other papers covered the following topics: the physical and chemical properties of dust in the coma of 67P/C-G based on Rosetta/COSIMA's in-situ measurements by Martin Hilchenbach \citep{hornung-et-al2016}; the dynamical properties of dust in the coma of 67P/C-G based on Rosetta/GIADA's in-situ measurements by Marco Fulle; a report on the latest results from the Rosetta/MIDAS experiment by Mark S. Bentley; an experimental study on the fragmentation, penetration, and consolidation of small bodies impacting on an asteroid by Akiko M. Nakamura; low-velocity impact experiments onto granular materials under reduced gravity conditions by Masato Kiuchi; a laboratory study on the infrared spectra of aggregate particles at a synchrotron facility by Akemi Tamanai; three-dimensional molecular dynamics simulations of polydisperse charged dust particles in a gravitational field by Robert Botet; far-infrared AKARI observations of asteroidal dust bands in the zodiacal light by Takafumki Ootsubo; an improvement on the COBE/DIRBE model of zodiacal emission based on the AKARI mid-infrared all-sky maps by Toru Kondo; the characterization of regolith dust particles using the wavelength dependence of the opposition effect of airless bodies by Ludmilla Kolokolova; an analysis of the COBE/DIRBE, AKARI, and MIRIS data for the zodiacal light toward the ecliptic poles by Jeonghyun Pyo; the determination of the aliphatic-to-aromatic carbon abundance ratio for Solar System objects from their near-infrared spectra by Li Zhou; a numerical study on the contribution of high-energy photoelectrons to dust levitation above the surface of airless bodies by Hiroki Senshu.

\subsection{Interstellar medium}

The Local Interstellar Cloud (LIC) is a partially ionized interstellar cloud that moves toward the galactic anticenter and currently surrounds the Sun \citep{bertin-et-al1993}.
The chemical composition of interstellar grains could be inferred from the argument of missing atoms, which is a classical method based on the depletions, namely, the underabundances of elements in the gas phase with respect to the cosmic abundances \citep[e.g.,][]{spitzer1978,whittet2003}.
A clear correlation between silicon and magnesium in the LIC and no correlation between silicon and iron indicate that interstellar grains are mainly composed of magnesium-rich silicates and iron-bearing alloys, as well as organic refractory materials \citep{kimura2015}.
In the past two decades, space missions have provided us the opportunity of studying the compositions of LIC grains that penetrate deep into the Solar System.
Andrew Westphal reported on the analyses of seven particles that were captured by the Stardust Interstellar Dust Collector, returned to Earth, and identified to be of LIC origin \citep{westphal-et-al2014,westphal2015}.
A typical picture of the interstellar dust candidates is magnesium-rich silicate grains with iron-rich phases without carbonaceous materials.
This is consistent with the mineralogical characteristics of interstellar grains derived from analyses of their chemical compositions measured in situ by Cassini near the Saturnian orbit \citep{altobelli-et-al2016}.
However, the organic refractory component of interstellar dust has likely been dissipated en route to the Solar System as the gas-to-dust mass ratio and the composition of pick-up ions have preserved a trace of missing organic matter \citep{kimura-et-al2015b}.
It is worth mentioning that porous interstellar grains with mass $m = 3$--$4\times {10}^{-15}\,\mathrm{kg}$ were identified by Stardust, while non-porous interstellar grains with $m < 5\times {10}^{-16}\,\mathrm{kg}$ were detected by Cassini.
In addition, a comparison of the Ulysses in-situ data on interstellar grains and numerical simulations of their dynamics in the Solar System suggests that the grains with $m > 2\times {10}^{-16}\,\mathrm{kg}$ are porous \citep{sterken-et-al2015}.
{\sl To achieve a consensus on this issue, small interstellar grains of $m \le 2\times {10}^{-16}\,\mathrm{kg}$ must be compact particles, while large interstellar grains of $m > 2\times {10}^{-16}\,\mathrm{kg}$ may be fluffy aggregates consisting of the small compact particles.}

Other papers covered the following topics: a review on the diffuse interstellar bands by Hongbin Ding; a report on the sublimation of organic material in interstellar dust en route to the Solar System by Hiroshi Kimura; a dynamical modeling of interstellar dust streaming into the Solar System by Veerle Sterken; numerical simulation on the mid-infrared extinction curve with micrometer-sized graphite interstellar grains by Shu Wang; a Monte Carlo modeling of UV radiation scattered at interstellar dust based on the GALEX data by Jayant Murthy; condensation experiments of silicate and carbon grains at cryogenic temperatures by Ga\"{e}l Rouill\'{e}; a modeling of dust properties in the Magellanic clouds based on Herschel and Spitzer observations by Fr\'{e}d\'{e}ric Galliano; infrared spectroscopic observations of ice absorption bands in the Large Magellanic Cloud by Takashi Shimonishi; a summary of AKARI's near-infrared spectroscopic data for interstellar ices by Takashi Onaka; a correlational study between far ultraviolet and mid infrared intensities in the Large Magellanic Cloud by Shalima Puthiyaveettil \citep{saikia-et-al2016}; a numerical study on the contribution of graphene to the interstellar extinction curve and infrared emission spectra by Xiuhui Chen; a demonstration of the AKARI far-infrared all-sky atlas by Yasuo Doi; near-infrared high resolution spectral observations of diffuse interstellar bands by Satoshi Hamano; data analyses of AKARI and ISO spectra from PAHs by Mark Hammonds; numerical simulation on infrared emission features of PAHs with various aliphatic substituents by Xuejuan Yang; a theoretical study on infrared spectral emission features of deuterated PAHs by Mridusmita Buragohain \citep{buragohain-et-al2016}.

\section{Perspectives for the development of cosmic dust research}
\label{perspectives}

We sometimes find that it is very difficult, if not impossible, to reach a consensus among experts even on a single issue of cosmic dust research.
Such a difficulty is clearly exemplified by a wide variety of models that have been proposed for interstellar dust over recent decades.
In {\sc Cosmic Dust \MakeUppercase{\romannumeral 6}}, Ant Jones introduced his new interstellar dust model (hereafter AJ) that reproduces the majority of observations and fulfills observational constraints \citep[see][]{jones-et-al2013}.
However, almost all the papers on interstellar dust presented two years later in {\sc Cosmic Dust \MakeUppercase{\romannumeral 8}} still rely on the classical model of \citet{draine-lee1984} (hereafter DL), in which interstellar dust is composed of either amorphous silicate (Mg$_{1.1}$Fe$_{0.9}$SiO$_4$) or graphite as refinement to the so-called MRN model after \citet{mathis-et-al1977}.
This indicates that the DL model is more popular than the AJ model, but does not necessarily mean that the former is more realistic than the latter.
The popularity of the DL model most probably results from its easiness of use, since the size distribution is described as a mathematically simple power-law and the dielectric functions of grain materials are given in tabular form \citep[cf.][]{draine1985}.
The DL model has remained in an area of consensus on the composition of interstellar dust where amorphous silicate and graphite are seemingly responsible for the $9.7\,\mathrm{\mu m}$ and $18\,\mathrm{\mu m}$ absorption features, and the $217.5\,\mathrm{nm}$ extinction feature, respectively.
However, the DL model may be incompatible with the latest findings of grain materials and sizes in the local interstellar medium, most of which have been presented in a series of Cosmic Dust meetings:
The size distribution of interstellar grains is skewed toward heavier masses by more than one order of magnitude and has a shallower power-law index than expected from the DL model \citep{kimura-et-al2003,sterken-et-al2015};
There is no evidence for the presence of carbonaceous materials in interstellar grains measured by Cassini and Stardust, although graphite grains, if exist as in the DL model, should have been identified by Cassini \citep{kimura-et-al2015b,westphal2015};
Interstellar grains extracted from the aerogel collector of Stardust contain forsteritic olivine, contrary to amorphous silicate expected from the DL model \citep{westphal2015,westphal-et-al2014}.
Very recently, \citet{wright-et-al2016} have found evidence for crystalline silicates in the interstellar medium from their infrared spectral observations toward the Galactic Center, in contrast to the DL model.
Furthermore, it turned out that interstellar grains are compositionally homogeneous and seem to contain magnesium-rich silicates and metallic iron, dissimilar from Mg$_{1.1}$Fe$_{0.9}$SiO$_4$ in the DL model \citep{altobelli-et-al2016}.
These recent discoveries of interstellar grain properties would give an impact on existing models for interstellar dust and will shed new light on unification between different models by consensus in the relevant scientific community.

Our understanding of the physical processes in the interstellar medium and in circumstellar environments is to some extent based on dust grain models under different conditions. 
The knowledge of these processes helps us to correctly interpret the spectra of astronomical objects and to deduce their compositions and structures from the correct interpretations. 
However, there are still many gaps in our understanding of the physical processes occurring on cosmic dust that have to be filled before we can finally state that dust grain models are completely realistic. 
Fortunately, we are living in an era of new and exciting observational instruments and techniques such as the space observatories Planck and Herschel, the Stratospheric Observatory for Infrared Astronomy (SOFIA), and the Atacama Large Millimeter Array (ALMA) that will reach unprecedented spatial resolutions and sensitivities. 
Astronomical observations of dust and molecular gas emission in circumstellar shells, molecular clouds, protoplanetary disks, and other dusty regions with the Herschel Space Telescope open new perspectives on the evolution of matter in the Universe. 
Consequently, many discoveries are still to be expected from a combination of astronomical observations, theoretical modelings, and laboratory studies in the next future. 
We cannot help but anticipate that this Cosmic Dust meeting series will act as a catalyst for new findings out of a comprehensive study on the evolution of cosmic dust.

The reader is cordially invited to take part in the Cosmic Dust meeting series and to join us for the development of cosmic dust research\footnote{Contact:~dust-inquries@cps-jp.org}.
The 9th meeting on Cosmic Dust ({\sc Cosmic Dust \MakeUppercase{\romannumeral 9}}) will be held at Tohuku University, Sendai, Japan on August 15--19, 2016.
More information on the Cosmic Dust meeting series is available at the Cosmic Dust website (https://www.cps-jp.org/\~{}dust).

\section*{Acknowledgements}
We express our sincere gratitude to Koji Wada (LOC Chair), Hiroki Senshu, Hiroki Chihara, Hiroshi Kobayashi, Takaya Nozawa, Takashi Shimonishi, Takayuki Hirai, Ryo Tazaki, and Akio Inoue for their cheerful dedication to the organization of {\sc Cosmic Dust \MakeUppercase{\romannumeral 8}} as a member of the LOC.
We thank all the authors and the reviewers as well as the editorial board of PSS and Elsevier for putting their efforts into this special issue. 
We are indebted to Osaka Sangyo University, JSPS (Japan Society for the Promotion of Science), National Astronomical Observatory of Japan (NAOJ), and Chiba Institute of Technology for their various supports to the organization of the meeting.
We would also like to acknowledge the grant of Research Assembly supported by the Research Coordination Committee of NAOJ, National Institutes of Natural Sciences.

\bibliographystyle{model2-names}
\bibliography{<your-bib-database>}

\begin{thebibliography}{00}

%% \bibitem must have one of the following forms:
%%   \bibitem[Jones et al.(1990)]{key}...
%%   \bibitem[Jones et al.(1990)Jones, Baker, and Williams]{key}...
%%   \bibitem[Jones et al., 1990]{key}...
%%   \bibitem[\protect\citeauthoryear{Jones, Baker, and Williams}{Jones
%%       et al.}{1990}]{key}...
%%   \bibitem[\protect\citeauthoryear{Jones et al.}{1990}]{key}...
%%   \bibitem[\protect\astroncite{Jones et al.}{1990}]{key}...
%%   \bibitem[\protect\citename{Jones et al., }1990]{key}...
%%   \harvarditem[Jones et al.]{Jones, Baker, and Williams}{1990}{key}...
%%

\bibitem[Aller et al.(2012)]{aller-et-al2012}
Aller, M.C., Kulkarni, V.P., York, D.G., Vladilo, G., Welty, D.E., Som, D., 2012. Interstellar silicate dust in the $z = 0.89$ absorber toward PKS 1830-211: crystalline silicates at high redshift? Astrophys. J. 748, 19(35pp). doi:10.1088/0004-637X/748/1/19

\bibitem[Aller et al.(2014)]{aller-et-al2014}
Aller, M.C., Kulkarni, V.P., York, D.G., Welty, D.E., Vladilo, G., Liger, N., 2014. Interstellar silicate dust in the $z = 0.685$ absorber toward TXS 0218+357. Astrophys. J. 785, 36(16pp). doi:10.1088/0004-637X/785/1/36

\bibitem[Altobelli et al.(2016)]{altobelli-et-al2016}
Altobelli, N., Postberg, F., Fiege, K., Trieloff, M., Kimura, H., Sterken, V.J., Hsu, H.-W., Hillier, J., Khawaya, N., Moragas-Klostermeyer, G., Blum, J., Burton, M., Srama, R., Kempf, S., Gruen, E., 2016, Flux and composition of interstellar dust at Saturn from Cassini's Cosmic Dust Analyzer. Science, 352, 312--318. doi:10.1126/science.aac6397

\bibitem[Aumann et al.(1984)]{aumann-et-al1984}
Aumann, H.H., Gillett, F.C., Beichman, C.A., de Jong, T., Houck, J.R., Low, F.J., Neugebauer, G., Walker, R.G., Wesselius, P.R., 1984. Discovery of a shell around Alpha Lyrae. Astrophys. J. 278, L23--L27. doi:10.1086/184214

\bibitem[Bertin et al.(1993)]{bertin-et-al1993}
Bertin, P., Lallement, R., Ferlet, R., Vidal-Madjar, A., 1993. Detection of the local interstellar cloud from high-resolution spectroscopy of nearby stars: Inferences on the heliospheric interface. J. Geophys. Res. 98, 15193--15197. doi:10.1029/93JA01179

\bibitem[Brownlee et al.(2005)]{brownlee-et-al2005}
Brownlee, D.E., Joswiak, D.J., Bradley, J.P., Matrajt, G., Wooden, D.H., 2005. Cooked GEMS -- Insight into the hot origin of crystalline silicates in circumstellar disks and the cold origin GEMS. Lunar Planet. Sci. XXXVI, 2391.

\bibitem[Buragohain et al.(2016)]{buragohain-et-al2016}
Buragohain, M., Pathak, A., Sarre, P., Onaka, T., Sakon, I., 2016. Mid-infrared vibrational study of deuterium-containing PAH variants. Planet. Sp. Sci. in this issue.

\bibitem[Draine(1985)]{draine1985}
Draine, B.T., 1985. Tabulated optical properties of graphite and silicate grains. Astrophys. J. Suppl. Ser. 57, 587--594. doi:10.1086/191016

\bibitem[Draine and Lee(1984)]{draine-lee1984}
Draine, B.T., Lee, H.M., 1984. Optical properties of interstellar graphite and silicate grains. Astrophys. J. 285, 89--108. doi:10.1086/162480

\bibitem[Dullemond(2013)]{dullemond2013}
Dullemond, C.P., 2013. Formation of (exo-)planets. Astron. Nachr. 334, 589--594. doi:10.1002/asna.201311899

\bibitem[Ertel et al.(2014)]{ertel-et-al2014}
Ertel, S., Absil, O., Defr\`{e}re, D., Le Bouquin, J.-B., Augereau, J.-C., Marion, L., Blind, N., Bonsor, A., Bryden, G., Lebreton, J., Milli, J., 2014. A near-infrared interferometric survey of debris-disk stars. IV. An unbiased sample of 92 southern stars observed in H band with VLTI/PIONIER. Astron. Astrophys. 570, A128. doi:10.1051/0004-6361/201424438

\bibitem[Flynn et al.(2013)]{flynn-et-al2013}
Flynn, G.J., Wirick, S., Keller, L. P., 2013. Organic grain coatings in primitive interplanetary dust particles: implications for grain sticking in the Solar Nebula. Earth Planets Sp. 65, 1159--1166. doi:10.5047/eps.2013.05.007

\bibitem[Granvik et al.(2016)]{granvik-et-al2016}
Granvik, M., Morbidelli, A., Jedicke, R., Bolin, B., Bottke, W.F., Beshore, E., Vokrouhlick\'{y}, D., Delb\`{o}, M., Michel, P., 2016. Super-catastrophic disruption of asteroids at small perihelion distances. Nature 530, 303--306. doi:10.1038/nature16934

\bibitem[Greenberg et al.(2000)]{greenberg-et-al2000}
Greenberg, J.M., Gillette, J.S., Mu\~{n}oz Caro, G.M., Mahajan, T.B., Zare, R.N., Li, A., Schutte, W.A., de Groot, M., Mendoza-G\'{o}mez, C., 2000. Ultraviolet photoprocessing of interstellar dust mantles as a source of polycyclic aromatic hydrocarbons and other conjugated molecules. Astrophys. J. 531, L71--L73. doi:10.1086/312526

\bibitem[He et al.(2014)]{he-et-al2014}
He, J., Jing, D., Vidali, G., 2014. Atomic oxygen diffusion on and desorption from amorphous silicate surfaces. Phys. Chem. Chem. Phys. 16, 3493--3500. doi:10.1039/c3cp54328e

\bibitem[He et al.(2015)]{he-et-al2015}
He, J., Shi, J., Hopkins, T., Vidali, G., Kaufman, M.J., 2015. A new determination of the binding energy of atomic oxygen on dust grain surfaces: experimental results and simulations. Astrophys. J. 801, 120(7pp). doi:10.1088/0004-637X/801/2/120

\bibitem[Hirashita et al.(2016)]{hirashita-et-al2016}
Hirashita, H., Nozawa, T., Asano, R.S., Lee, T., 2016. Revisiting the lifetime estimate of large presolar grains in the interstellar medium. Planet. Sp. Sci. in this issue.

\bibitem[Hooks et al.(2016)]{hooks-et-al2016}
Hooks, J.M., et al., 2016. Optimized dust velocity grid system aiding the compositional mapping of Jovian satellites. Planet. Sp. Sci. in this issue.

\bibitem[Hornung et al.(2016)]{hornung-et-al2016}
Hornung, K., Merouane, S., Hilchenbach, M., Langevin, Y., Mellado, E.M., Della Corte, V. Kissel, J., Engrand, C., Schulz, R., Ryno, J., Silen, J., the COSIMA team, 2016. A first assessment of the strength of cometary particles collected in-situ by the COSIMA instrument onboard ROSETTA. Planet. Sp. Sci. in this issue.

\bibitem[J\"{a}ger et al.(2003)]{jaeger-et-al2003}
J\"{a}ger, C., Fabian, D., Schrempel, F., Dorschner, J., Henning, T., Wesch, W., 2003. Structural processing of enstatite by ion bombardment. Astron. Astrophys. 401, 57--65. doi:10.1051/0004-6361:20030002

\bibitem[Jenkins(2009)]{jenkins2009}
Jenkins, E.B., 2009. A unified representation of gas-phase element depletions in the interstellar medium. Astrophys. J. 700, 1299--1348. doi:10.1088/0004-637X/700/2/1299

\bibitem[Jones et al.(2013)]{jones-et-al2013}
Jones, A.P., Fanciullo, L., K\"{o}hler, M., Verstraete, L., Guillet, V., Bocchio, M., Ysard, N., 2013. The evolution of amorphous hydrocarbons in the ISM: dust modelling from a new vantage point. Astron. Astrophys. 558, A62. doi:10.1051/0004-6361/201321686

\bibitem[Keller et al.(2000)]{keller-et-al2000}
Keller, L.P., Messenger, S., Bradley, J.P., 2000. Analysis of a deuterium-rich interplanetary dust particle (IDP) and implications for presolar material in IDPs. J. Geophys. Res. 105, 10397--10402. doi:10.1029/1999JA900395

\bibitem[Kennedy(2015)]{kennedy2015} 
Kennedy, G., 2015. Debris Disks: Seeing Dust, Thinking of Asteroids, Comets, and Planets. Presented at {\sc Cosmic Dust \MakeUppercase{\romannumeral 8}}. $\langle$https://www.cps-jp.org/\~{}dust/Program\_VIII\_files/dust8-2-4-1.pdf$\rangle$

\bibitem[Kimber et al.(2014)]{kimber-et-al2014}
Kimber, H.J., Ennis, C.P., Price, S.D., 2014. Single and double addition of oxygen atoms to propyne on surfaces at low temperatures. Faraday Discuss. 168, 167--184. doi:10.1039/C3FD00130J

\bibitem[Kimura(2015)]{kimura2015}
Kimura, H., 2015. Interstellar dust in the Local Cloud surrounding the Sun. Mon. Not. R. Astron. Soc. 449, 2250--2258. doi:10.1093/mnras/stv427

\bibitem[Kimura and Mann(1998)]{kimura-mann1998}
Kimura, H., Mann, I., 1998. Brightness of the solar F-corona. Earth Planets Sp. 50, 493--499. doi:10.1186/BF03352140

\bibitem[Kimura et al.(2003)]{kimura-et-al2003}
Kimura, H., Mann, I., Jessberger, E.K., 2003. Composition, structure, and size distribution of dust in the Local Interstellar Cloud. Astrophys. J. 583, 314--321. doi:10.1086/345102

\bibitem[Kimura et al.(2015a)]{kimura-et-al2015a}
Kimura, H., Wada, K., Senshu, H., Kobayashi, H., 2015a. Cohesion of amorphous silica spheres: toward a better understanding of the coagulation growth of silicate dust aggregates. Astrophys. J. 812, 67(12pp). doi:10.1088/0004-637X/812/1/67

\bibitem[Kimura et al.(2015b)]{kimura-et-al2015b}
Kimura, H., Postberg, F., Altobelli, N., Trieloff, M., 2015b. Missing Organic Materials From Interstellar Dust Inside the Solar System. Presented at {\sc Cosmic Dust \MakeUppercase{\romannumeral 8}}. $\langle$https://www.cps-jp.org/\~{}dust/Program\_VIII\_files/dust8-5-1-2.pdf$\rangle$

\bibitem[Kolokolova et al.(2016)]{kolokolova-et-al2016}
Kolokolova, L., Nagdimunov, L., A'Hearn, M., King, A., Wolff, M., 2016. Studying the nucleus of Comet 9P/Tempel 1 using the structure of the Deep Impact ejecta cloud at the early stages of its development. Planet. Sp. Sci. in this issue.

\bibitem[Kr\"{a}tschmer and Huffman(1979)]{kraetschmer-huffman1979}
Kr\"{a}tschmer, W., Huffman, D.R., 1979. Infrared extinction of heavy ion irradiated and amorphous olivine, with applications to interstellar dust. Astrophys. Sp. Sci. 61, 195--203. doi:10.1007/BF00645803

\bibitem[Kr\"{u}ger et al.(2015)]{krueger-et-al2015} 
Kr\"{u}ger, H., Seidensticker, K.J., Fischer, H.-H., Albin, T., Apathy, I., Arnold, W., Flandes, A., Hirn, A., Kobayashi, M., Loose, A., P\'{e}ter, A., Podolak, M., 2015. Dust Impact Monitor (SESAME-DIM) measurements at comet 67P/Churyumov-Gerasimenko. Astron. Astrophys. 583, A15. doi:10.1051/0004-6361/201526400

\bibitem[Kulkarni et al.(2016)]{kulkarni-et-al2016} 
Kulkarni, V.P., Aller, M.C., York, D.G., Welty, D.E., Vladilo, G., Som, D., 2016. Probing the interstellar dust in galaxies over $> 10$~Gyr of cosmic history. Planet. Sp. Sci. in this issue.

\bibitem[Marboeuf et al.(2016)]{marboeuf-et-al2016} 
Marboeuf, U., Bonsor, A., Augereau, J.-C., 2016. Extrasolar comets: the origin of dust in exozodiacal disks? Planet. Sp. Sci. in this issue.

\bibitem[Mathis et al.(1977)]{mathis-et-al1977} 
Mathis, J.S., Rumpl, W., Nordsieck, K.H., 1977. The size distribution of interstellar grains. Astrophys. J. 217, 425--433. doi:10.1086/155591

\bibitem[Mattsson(2016)]{mattsson2016} 
Mattsson, L., 2016. Modelling dust processing and the evolution of grain sizes in the ISM using the method of moments. Planet. Sp. Sci. in this issue.

\bibitem[Mauney and Lazzati(2016)]{mauney-lazzati2016}
Mauney, C.M., Lazzati, D., 2016. Formation and properties of astrophysical carbonaceous dust. Planet. Sp. Sci. in this issue.

\bibitem[Minissale et al.(2016)]{minissale-et-al2016} 
Minissale, M., Congiu, E., Dulieu, F., 2016. Direct measurement of desorption and diffusion energies of O and N atoms physisorbed on amorphous surfaces. Astron. Astrophys. 585, A146. doi:10.1051/0004-6361/201526702

\bibitem[Molster and Kemper(2005)]{molster-kemper2005}
Molster, F., Kemper, C., 2005. Crystalline silicates. Sp. Sci. Rev. 119, 3--28. doi:10.1007/s11214-005-8066-x

\bibitem[Nozawa(2016)]{nozawa2016}
Nozawa, T., 2016. Properties of interstellar dust responsible for extinction laws with unusually low total-to-selective extinction ratios of $R_\mathrm{V} = 1$--$2$. Planet. Sp. Sci. in this issue.

\bibitem[Pinte(2015)]{pinte2015} 
Pinte, C., 2015. Dust Evolution in Protoplanetary Disks. Presented at {\sc Cosmic Dust \MakeUppercase{\romannumeral 8}}. $\langle$https://www.cps-jp.org/\~{}dust/Program\_VIII\_files/dust8-2-2-1.pdf$\rangle$

\bibitem[Podolak et al.(2016)]{podolak-et-al2016}
Podolak, M., Flandes, A., Della Corte, V., Kr\"{u}ger, H., 2016. A simple model for understanding the DIM dust measurement at comet 67P/Churyumov-Gerasimenko. Planet. Sp. Sci. in this issue.

\bibitem[Poteet et al.(2015)]{poteet-et-al2015}
Poteet, C.A., Whittet, D.C.B., Draine, B.T., 2015. The composition of interstellar grains toward $\zeta$ Ophiuchi: constraining the elemental budget near the diffuse-dense cloud transition. Astrophys. J. 801, 110(12pp). doi:10.1088/0004-637X/801/2/110

\bibitem[Saikia et al.(2016)]{saikia-et-al2016}
Saikia, G., Shalima, P., Gogoi, R., Pathak, A., 2016. Comparison of diffuse infrared and far-ultraviolet emission in the Large Magellanic Cloud: the data. Planet. Sp. Sci. in this issue.

\bibitem[Smith and Terrile(1984)]{smith-terrile1984}
Smith, B.A., Terrile, R.J., 1984. A circumstellar disk around $\beta$ Pictoris. Science 226, 1421--1424. doi:10.1126/science.226.4681.1421

\bibitem[Spitzer(1978)]{spitzer1978} 
Spitzer, L., 1978. Physical Processes in the Interstellar Medium. Wiley-VCH, Weinheim. Section~1.41.

\bibitem[Sterken et al.(2015)]{sterken-et-al2015}
Sterken, V.J., Strub, P., Kr\"{u}ger, H., von Steiger, R., Frisch, P., 2015. Sixteen years of Ulysses interstellar dust measurements in the Solar System. III. Simulations and data unveil new insights into local interstellar dust. Astrophys. J. 812, 141(24pp). doi:10.1088/0004-637X/812/2/141

\bibitem[Taylor et al.(2015)]{taylor-et-al2015}
Taylor, M.G.G.T., Alexander, C., Altobelli, N., Fulle, M., Fulchignoni, M., Grün, E., Weissman, P., 2015. Rosetta begins its comet tale. Science 347, 387--387. doi:10.1126/science.aaa4542

\bibitem[Tazaki et al.(2015)]{tazaki-et-al2015}
Tazaki, R., Okuzumi, S., Kataoka, A., Tanaka, H., Nomura, H., 2015. Optical Properties of Fractal Dust Aggregates. Presented at {\sc Cosmic Dust \MakeUppercase{\romannumeral 8}}. $\langle$https://www.cps-jp.org/\~{}dust/Program\_VIII\_files/dust8-P06.pdf$\rangle$

\bibitem[Vidali(2015)]{vidali2015} 
Vidali, G., He, J., Hopkins, T., Shi, J., Kaufman, M., 2015. Measurement of Binding of Oxygen on Dust Grains (Amorphous Silicates and Amorphous Water Ice): Consequences for Oxygen Chemistry in the Interstellar Medium. Presented at {\sc Cosmic Dust \MakeUppercase{\romannumeral 8}}. $\langle$https://www.cps-jp.org/\~{}dust/Program\_VIII\_files/dust8-2-1-1.pdf$\rangle$

\bibitem[Wada et al.(2009)]{wada-et-al2009}
Wada, K., Tanaka, H., Suyama, T., Kimura, H., Yamamoto, T., 2009. Collisional growth conditions for dust aggregates. Astrophys. J. 702, 1490--1501. doi:10.1088/0004-637X/702/2/1490

\bibitem[Wang et al.(2014)]{wang-et-al2014} 
Wang, S., Li, A., Jiang, B.W., 2014. Modeling the infrared interstellar extinction. Planet. Space Sc. 100, 32--39. doi:10.1016/j.pss.2014.03.018

\bibitem[Wang et al.(2015)]{wang-et-al2015} 
Wang, S., Li, A., Jiang, B.W., 2015. The interstellar oxygen crisis, or where have all the oxygen atoms gone? Mon. Not. R. Astron. Soc. 454, 569--575. doi:10.1093/mnras/stv1900

\bibitem[Ward and Price(2011)]{ward-price2011}
Ward, M.D., Price, S.D., 2011. Thermal reactions of oxygen atoms with alkenes at low temperatures on interstellar dust. Astrophys. J. 741, 121(9pp). doi:10.1088/0004-637X/741/2/121

\bibitem[Weidenschilling et al.(1989)]{weidenschilling-et-al1989}
Weidenschilling, S., Donn, B., Meakin, P., 1989. The physics of planetesimal formation. In: Weaver, H., Danly, L., Fall, S. (Eds.), The Formation and Evolution of Planetary Systems. Cambridge Univ. Press, Cambridge, pp. 131--150.

\bibitem[Westphal et al.(2014)]{westphal-et-al2014}
Westphal, A.J., Stroud, R.M., Bechtel, H.A., Brenker, F.E., Butterworth, A.L., Flynn, G.J., et al., 2014. Evidence for interstellar origin of seven dust particles collected by the Stardust spacecraft. Science 345, 786--791. doi:10.1126/science.1252496

\bibitem[Westphal(2015)]{westphal2015}
Westphal, A.J., 2015. Laboratory Analyses of Seven Particles of Likely Interstellar Origin Returned by the Stardust Spacecraft. Presented at {\sc Cosmic Dust \MakeUppercase{\romannumeral 8}}. $\langle$https://www.cps-jp.org/\~{}dust/Program\_VIII\_files/dust8-5-1-1.pdf$\rangle$

\bibitem[Whittet(2003)]{whittet2003} 
Whittet, D.C.B., 2003. Dust in the Galactic Environment, 2nd ed. Institute of Physics, Bristol. Section~2.4.

\bibitem[Whittet(2010)]{whittet2010} 
Whittet, D.C.B., 2010. Oxygen depletion in the interstellar medium: implications for grain models and the distribution of elemental oxygen. Astrophys. J. 710, 1009--1016. doi:10.1088/0004-637X/710/2/1009

\bibitem[Whittet et al.(1988)]{whittet-et-al1988} 
Whittet, D.C.B., Bode, M.F., Longmore, A.J., Adamson, A.J., McFadzean, A.D., Aitken, D.K., Roche, P.F., 1988. Infrared spectroscopy of dust in the Taurus dark clouds: ice and silicates. Mon. Not. R. Astron. Soc. 233, 321--336. doi:10.1093/mnras/233.2.321

\bibitem[Wright et al.(2016)]{wright-et-al2016}
Wright, C.M., Do Duy, T., Lawson, W., 2016. Absorption at $11\,\mathrm{\mu m}$ in the interstellar medium and
embedded sources: evidence for crystalline silicates. Mon. Not. R. Astron. Soc. 457, 1593--1625. doi:10.1093/mnras/stw041

\bibitem[Xie et al.(2016)]{xie-et-al2016}
Xie, Y., Nikutta, R., Hao, L. Li, A., 2016. A tale of three galaxies: a ``CLUMPY'' view of the spectroscopically anomalous galaxies IRAS F10398+1455, IRAS F21013-0739 and SDSS J0808+3948. Planet. Sp. Sci. in this issue.

\bibitem[Zafar(2016)]{zafar2016}
Zafar, T., 2016. GRB afterglows: dust extinction properties from the low to high redshift universe. Planet. Sp. Sci. in this issue.


\end{thebibliography}

%% not want to use model2-names.bst.


\begin{flushright}
Hiroshi Kimura\\
{\sl Graduate School of Science, Kobe University, c/o CPS (Center for Planetary Science), Chuo-ku Minatojima Minamimachi 7-1-48, Kobe 650-0047, Japan}

\vspace{0.2in}
Ludmilla Kolokolova\\
{\sl Planetary Data System Group, Department of Astronomy, University of Maryland, College Park, MD 20742, USA}

\vspace{0.2in}
Aigen Li\\
{\sl Department of Physics and Astronomy, University of Missouri, 314 Physics Building, Columbia, MO 65211, USA}

\vspace{0.2in}
Hidehiro Kaneda\\
{\sl Graduate School of Science, Nagoya University, Furo-cho, Chikusa-ku, Nagoya 464-8602, Japan}

\vspace{0.2in}
Cornelia J\"{a}ger\\
{\sl Max Planck Institute for Astronomy, Heidelberg, Laboratory Astrophysics and Clusterphysics Group at the Institute of Solid State Physics, Friedrich Schiller University Jena, Helmholtzweg 3, 07743 Jena, Germany}

\vspace{0.2in}
Jean-Charles Augereau\\
{\sl Universit\'e Grenoble Alpes, IPAG, F-38000 Grenoble, France}\\
{\sl CNRS, IPAG, F-38000 Grenoble, France}
\end{flushright}

%\end{linenumbers}
\end{document}